\documentclass[a4paper]{jpconf}
\usepackage{graphicx}
\usepackage{bbm}
\usepackage{amsfonts}
\usepackage{amsthm}
\usepackage{amssymb}
\usepackage{amsmath}

\newcommand{\beq}{\begin{eqnarray}}
\newcommand{\eeq}{\end{eqnarray}}
\newcommand{\ist}{\;=\;}

\begin{document}
\begin{flushright}UWThPh-2011-11\end{flushright}
\title{
The Entropic Uncertainty Principle for Decaying Systems and $\mathcal{CP}$ violation}

\author{Beatrix C. Hiesmayr}

\address{University of Vienna, Faculty of Physics, Boltzmanngasse 5, 1090 Vienna, Austria; \textit{and}\\
Research Center for Quantum Information, Institute of Physics,
Slovak Academy of Sciences, D\'ubravsk\'a cesta 9, 84511 Bratislava, Slovakia}

\ead{Beatrix.Hiesmayr@univie.ac.at}

\begin{abstract}
Employing an effective formalism for decaying system we are able to investigate Heisenberg's uncertainty relation for observables measured at accelerator facilities. In particular we investigate the neutral K--meson system and show that, firstly, due to the time evolution an uncertainty between strangeness measurements at different times is introduced and, secondly, due to the imbalance of matter and antimatter ($\mathcal{CP}$ violation) an uncertainty in the evolution of the eigenstates of the effective Hamiltonian of the system. Consequently, the existence of $\mathcal{CP}$ violation is linked to uncertainties of observables, i.e. the outcomes cannot be predicted even \textit{in principle} to arbitrary precisions.
\end{abstract}

% Title, authors and addresses

% use the thanksref command within \title, \author or \address for footnotes;
% use the corauthref command within \author for corresponding author footnotes;
% use the ead command for the email address,
% and the form \ead[url] for the home page:

%%%%%%%%%%%%%%%%%%%%%%%%%%%%%%%%%%%%%%%%%%%%%%%%%%%%%%%%%%%%%%%%%%%%%%%
\section{Introduction}
%%%%%%%%%%%%%%%%%%%%%%%%%%%%%%%%%%%%%%%%%%%%%%%%%%%%%%%%%%%%%%%%%%%%%%%
Heisenberg's uncertainty relation is one of the most well known phenomenon in popular science, it captures one of the key features of quantum theory and essentially describes the stability of atoms. In detail it bounds the uncertainties of the outcomes of any two observables on a system in terms of their commutativity or differently stated uncertainty relations constrain the potential knowledge an observer can have  about the physical properties of a system. In high energy physics one usually observes different decay events at different times corresponding to different distances from the point of particle creation. In this note we enlighten the uncertainties in the time evolution of unstable two--state systems with two different decay constants.

We start by repeating Heisenberg's uncertainty relation and presenting an entropic version of it. Considering typical observables in the neutral K-meson system including $\mathcal{CP}$ violation (${\cal C}\dots$charge conjugation; ${\cal P}\dots$ parity) we show the uncertainties in this system. This symmetry violation,  which captures the difference between a world of matter and a world of antimatter, has been proven in many precision experiments but its origin is still an open puzzle in Particle Physics. In this note we show that it adds uncertainty to the system. Deriving the uncertainty relation independent of the initial state of the system is only possible because we use an effective formalism which enables us to rewrite any observable in the Heisenberg picture, i.e. time--dependent, and effectively living in the surviving part of the in general infinite dimensional Hilbert space which was recently introduced in Ref.~\cite{Heisikaon}.

Meson--antimeson systems, as e.g. the neutral K--meson or B--meson system, have turned out to be very suitable to discuss various quantum foundation issues as
e.g. quantum erasing and marking procedures \cite{Hiesmayr7,Hiesmayr8, Hiesmayr10, HiesmayrKLOE} or Bell inequalities \cite{Hiesmayr1,Hiesmayr2,Hiesmayr3,Hiesmayr13,Bramon3,Bramon4,LIQiao} or for testing decoherence or entanglement in the system \cite{Hiesmayr6,Mavromatos1,Mavromatos2,Giuseppe1,Blasone,Giuseppe2,Hiesmayr5,Caban,Yabsley} or Bohr's complementary relation \cite{Hiesmayr9} or possible $\mathcal{CPT}$ violation effects \cite{Mavromatos3,DiDomenico,Capolupo1}.

%%%%%%%%%%%%%%%%%%%%%%%%%%%%%%%%%%%%%%%%%%%%%%%%%%%%%%%%%%%%%%%%%%%%%%%
\section{The Uncertainty Principle For Two Observables}
%%%%%%%%%%%%%%%%%%%%%%%%%%%%%%%%%%%%%%%%%%%%%%%%%%%%%%%%%%%%%%%%%%%%%%%

Let us repeat the derivation of the Robertson version \cite{Robertson} of the uncertainty principle of Heisenberg for two arbitrary observables $O_n, O_m$. Consider an arbitrary operator $C$ and note that
\beq
\langle C C^\dagger\rangle_\psi&=&\langle C^\dagger C\rangle_\psi\;\geq\; 0
\eeq
as $C C^\dagger$ is Hermitian. Consider two Hermitian operators $O_n, O_m$ with $(\Delta O)_\psi^2:=\langle O^2\rangle_\psi-\langle O\rangle_\psi^2$ being the standard mean value and choose $C\ist\frac{O_n-\langle O_n\rangle}{\Delta O_n}+i \frac{O_m-\langle
O_m\rangle}{\Delta O_m}$  then it follows that the inequality  \beq\label{Robertsonversion} (\Delta
O_n)_\psi\cdot(\Delta
O_m)_\psi\geq\frac{1}{2}\left|\langle[O_n,O_m]\rangle_\psi\right|\eeq has to be satisfied for any state $\psi$.
In case of $O_n=\hat{x}$ and $O_m=\hat{p}$, being the position and momentum operator, we obtain Heisenberg's well known uncertainty relation
$\Delta\hat{x}\Delta\hat{p}\;\geq\;\frac{\hbar}{2}$.

It was criticized that the version of the uncertainty principle by Robertson~(\ref{Robertsonversion}) has a lower bound which is depending in general on the state. It was shown that there exists an information theoretic formulation of uncertainty principle which lacks this problem and has in general stronger bounds~\cite{Iwo}, hence puts a stronger limit on the extent to which the two observables can be simultaneously peaked. The entropic uncertainty relation of two non-degenerate observables is given by (introduced by D. Deutsch \cite{Deutsch}, improved in Ref.~\cite{Kraus} and proven in Ref.~\cite{MaassenUffink})
\begin{eqnarray}\label{EQUI}
H(O_n)+H(O_m)&\geq& %2 \log_2\left(\frac{1}{\max_{i,j}\{|\langle\chi_n^i|\chi_m^j\rangle|\}}\right)\nonumber\\&=&
- 2\log_2\left(\max_{i,j}\{|\langle\chi_n^i|\chi_m^j\rangle|\}\right)
\end{eqnarray}
where  \begin{eqnarray}
H(O_n)=-p(n)\log_2 p(n)-(1-p(n))\log_2 (1-p(n))
\end{eqnarray}
is the binary entropy for a certain prepared pure state $\psi$ and the $p(n)$'s are the probability distribution associated with the measurement of $O_n$ for $\psi$,  hence $p(n)=|\langle \chi_n|\psi\rangle|^2$. The maximal value of the right hand side of the entropic uncertainty relation is obtained for
\begin{eqnarray}
|\langle \chi_n|\chi_m\rangle|&=&\frac{1}{\sqrt{2}}\;,
\end{eqnarray}
in this case the two observables are commonly called complementary to each other (their eigenvalues have to be nondegenerate), for example if the operators are $\sigma_x$ and $\sigma_z$. In general a non-zero value of the right hand side of  inequality (\ref{EQUI}) means that the two observables do not commute, i.e. it quantifies the complementarity of the observables. The binary entropies on the left hand side quantify the gain of information on average when we learn about the value of the random variable associated to $O_n$. Alternatively, one can interpret the entropy as the uncertainty \textit{before} we obtain the result of the random variable. Remarkably, the right hand side of the entropic uncertainty relation also does not depend on the eigenvalues (except to test the non-degeneracy), this means that if the state is prepared in an eigenstate say of $O_n$ then the two eigenvalues of $O_m$ are equally probable as measured values, i.e. the exact knowledge of the measured value of one observable implies maximal uncertainty of the measured value of the other, independent of the eigenvalues.

%%%%%%%%%%%%%%%%%%%%%%%%%%%%%%%%%%%%%%%%%%%%%%%%%%%%%%%%%%%%%%%%%%%%%%%%%%%%%%%%%%%
\section{Observables at Accelerator Facilities in the Heisenberg Picture}
%%%%%%%%%%%%%%%%%%%%%%%%%%%%%%%%%%%%%%%%%%%%%%%%%%%%%%%%%%%%%%%%%%%%%%%%%%%%%%%%%%%%%%

Particle detectors at accelerator facilities detect or reconstruct different decay products at different distances from the point of creation, usually by a \textit{passive} measurement procedure, i.e. observing a certain decay channel at a certain position in the detector without having control over the decay channel nor the time point (distance from point of creation). More rarely by an \textit{active} measurement procedure, e.g. placing a block of matter in the beam and forcing the incoming neutral meson to react with the material. In such a way e.g. the strangeness content of a neutral kaon at a certain time point (position of the block of matter) can be measured \textit{actively}, i.e. the experimenter decides what physical property (strangeness) and when. For measurements of the lifetime of neutral kaons also an \textit{active} measurement procedure was found in Refs.~\cite{Hiesmayr7,Hiesmayr8}, one waits for $4.8 \tau_S$ ($\tau_S$ lifetime of the short lived state) and if one observes a decay (any decay channel) within this time duration one denotes the event as a short lifetime measurement otherwise as a long lifetime measurement. In this case we have a conceptual error which is of the amount of $\mathcal{CP}$ violation (for details consult Ref.~\cite{Hiesmayr8})

In principle one can generally ask the following questions to the mesonic quantum systems
\begin{itemize}
\item Are you a certain quasispin $|k_n\rangle$ at a certain time $t_n$ or not?
\item Or: Are you a certain quasispin $|k_n\rangle$ or its orthogonal state $|k_n^\perp\rangle$ ($\langle k_n^\perp|k_n\rangle=0$) at a certain time $t_n$?
\end{itemize}
where we denote by a quasispin $k_n$ any superposition of the mass eigenstates $|K_{S/L}\rangle$ which are the solutions of the effective Schr\"odinger equation, i.e. $|k_n\rangle\ist \cos\frac{\alpha}{2}|K_S\rangle+\sin\frac{\alpha}{2}\cdot e^{i \phi_n}\;|K_L\rangle$. We will consider the first question because it includes all information that is \textit{in principle} obtainable in these decaying systems.

Let us now discuss what is learnt by finding a certain quasispin $|k_n\rangle$ at a certain time $t_n$ or not, compared to the situation to find a quasispin $k_m$ or not at the point of creation $t_m$. Certainly, this result also quantifies our uncertainty \textit{before} we learn the result (Yes, No) at $t_n$ and (Yes, No) at $t_m$. In particular, if we compare observables of same quasispin at different times, we obtain the uncertainty due to the time evolution.

Differently stated, we can view it in the following way, two experimenters, Alice and Bob, choose two different measurements corresponding to the observables $O_n$ and $O_m$. Alice prepares a certain state $\psi$ and sends it to Bob. Bob carries out one of the two measurements  $O_n$ or $O_m$ and announces his choice $n$ or $m$ to Alice. She wants to minimize her uncertainty about Bob's measurement result. Alice's result is consequently bounded by the inequality~(\ref{EQUI}).

Usually, the effect of the time evolution is asserted to the state vector describing the quantum system to a given time point. The difficulty with unstable systems is that then the state vector is no longer normalized and a simple re-normalizing by dividing by the norm of the state vector neglects some information before that particular time point that is \textit{in principle} obtainable. A way out is to consider decaying systems as open quantum system and enlarging there Hilbert space by at least twice, see Ref.~\cite{BGH4}. Herewith, the effect of the decay of these unstable systems can be effectively described as a certain ``\textit{kind of decoherence}'' as it is effectively described by a Markovian master equation. The price to pay is that for $t>0$ the state is no longer pure and the time evolution of multipartite meson systems is complicated to compute. Another way out was recently presented in Ref.~\cite{Heisikaon} by using the Heisenberg picture, i.e. making the observables time dependent, and by that being able to effectively describe the system by a Hilbert space that has the same dimension as number of degrees of freedom.

%%%%%%%%%%%%%%%%%%%%%%%%%%%%%%%%%%%%%%%%%%%%%%%%%%%%%%%%%%%%%%%%%%%%%%%%%%%%%%%%%%%
\section{An Effective Formalism For Decaying Systems}
%%%%%%%%%%%%%%%%%%%%%%%%%%%%%%%%%%%%%%%%%%%%%%%%%%%%%%%%%%%%%%%%%%%%%%%%%%%%%%%%%%%%%%

For two-state decaying systems any expectation value for finding a certain quasispin  $|k_n\rangle$ at a certain time $t_n$ or not
for given initial state $\rho$ (not necessarily pure) is given by
\begin{eqnarray}
E(k_n,t_n)&=& P(\textrm{Yes}: k_n,t_n)-P(\textrm{No}: k_n,t_n)\nonumber\\
&\stackrel{P(\textrm{No}: k_n,t_n)+P(\textrm{Yes}: k_n,t_n)=1}{=}&2\; P(\textrm{Yes}: k_n,t_n)-1\nonumber\\
&=& Tr (O_n^{eff}\;\rho)
\end{eqnarray}
with the effective $2\times 2$ operator
\begin{eqnarray}
O_n^{eff}&:=&O^{eff}(\theta_n,\phi_n, t_n)\ist\lambda_1(\theta_n,t_n,\delta)\;|\chi_n^1\rangle\langle\chi_n^1|+\lambda_2\;|\chi_n^2\rangle\langle\chi_n^2|
\end{eqnarray}
with the eigenvectors
\begin{eqnarray}
|\chi_n^{1}\rangle&=&
\frac{1}{\sqrt{N(t_n)}}
\biggl\lbrace \langle K_S|k_n\rangle\cdot e^{i\lambda_S^* t_n}\;|K_1\rangle+\langle K_L|k_n\rangle\cdot e^{i\lambda_L^* t_n}\;|K_2\rangle\biggr\rbrace\nonumber\\
|\chi_n^{2}\rangle&=&
\frac{1}{\sqrt{N(-t_n)}}
\biggl\lbrace -\langle K_L|k_n\rangle^*\cdot e^{i\lambda_S t_n}\;|K_1\rangle+\langle K_S|k_n\rangle^*\cdot e^{i\lambda_L t_n}\;|K_2\rangle\biggr\rbrace\nonumber\\
\textrm{with}\quad
N(t_n)&=&e^{-\Gamma_S t_n}\; |\langle K_S|k_n\rangle|^2+e^{-\Gamma_L t_n}\; |\langle K_L|k_n\rangle|^2
\end{eqnarray}
and the eigenvalues
\begin{eqnarray}
\lambda_1(\theta_n,t_n,\delta)&=&-1+ (e^{-\Gamma_S t_n}-e^{-\Gamma_L t_n}) (1-\delta^2) \cos\theta_n\nonumber\\
&&+ (e^{-\Gamma_S t_n}+e^{-\Gamma_L t_n})(1+\delta^2+2\delta \cos\phi_n \sin\theta_n)\nonumber\\
\lambda_2&=&-1\;.
\end{eqnarray}
Here $\Gamma_{S,L}$ are the decay constants of the eigenvectors of the Hamiltonian, i.e. the short lived state $|K_S\rangle$ and the long lived state $|K_L\rangle$. $|K_{1/2}\rangle$ denote the $\mathcal{CP}$ eigenstates and $\delta$ is the $\mathcal{CP}$ violating parameter. Note that due to the decaying property of the system the first eigenvalue $\lambda_1$ changes in time and approaches for $t_n\longrightarrow\infty$ the value of the second eigenvalue $\lambda_2$ which is independent of any choice of the observer.

For spin--$\frac{1}{2}$ systems, the most general observable is given by $\vec{n}\vec{\sigma}$ ($\sigma_i$ Pauli operators) where any normalized quantization direction ($|\vec{n}|=1$) can be parameterized by the azimuth angle $\theta_n$ and the polar angle $\phi_n$. In case of unstable systems the effective observable is given by the set of operators $O_n^{eff}=(|\vec{n}(\theta_n,\phi_n, t_n)|-1)\mathbbm{1}+\vec{n}(\theta_n,\phi_n,t_n)\vec{\sigma}$ for which the ``\textit{quantization direction}'' is no longer normalized and its loss results in an additional contribution in form of ``white noise'', i.e. the expectation value gets a contribution independent of the initial state for $t_n>0$.

%%%%%%%%%%%%%%%%%%%%%%%%%%%%%%%%%%%%%%%%%%%%%%%%%%%%%%%%%%%%%%%%%%%%%%%%%%%%%%%%%%%
\section{The Entropic Uncertainty Principle in the Neutral Kaon Systems}
%%%%%%%%%%%%%%%%%%%%%%%%%%%%%%%%%%%%%%%%%%%%%%%%%%%%%%%%%%%%%%%%%%%%%%%%%%%%%%%%%%%%%%

Equipped with this effective observable formalism in the Heisenberg picture we can derive the lower bound on the entropic uncertainty relation (\ref{EQUI}), for which we have to find the maximum of the normalized eigenvectors of the effective operators $O_n^{eff}$ and $O_m^{eff}$
\begin{eqnarray}
&&\max\biggl\{|\langle\chi_m^1|\chi_n^1\rangle|,|\langle\chi_m^1|\chi_n^2\rangle|,|\langle\chi_m^2|\chi_n^1\rangle|,|\langle\chi_m^2|\chi_n^2\rangle|\biggr\}\;.
\end{eqnarray}

\begin{figure}
\centering
\includegraphics[scale=0.5]{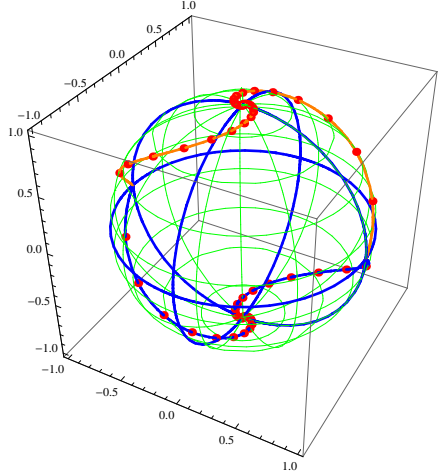}
%(b)\includegraphics[scale=0.5]{blochstrange.eps}
\caption{(Colour online) This figure depicts the Bloch sphere for the eigenvectors of observable $O^{eff}(\frac{\pi}{2},0,tn)$ ($\equiv|K^0\rangle$)                or $O^{eff}(\frac{\pi}{2},\pi,tn)$ ($\equiv|\bar K^0\rangle$) respectively. The eigenvectors $\chi^1$ represented by the corresponding Bloch vectors $\vec{b}^i_n$, Eq.~(\ref{Blochvector}), start at the equatorial plane and move with time (red dots) to the south pole ($\equiv|K_L\rangle$) while the eigenvectors $\chi^2$ being orthogonal to $\chi^1$ consequently move to the north pole ($\equiv|K_S\rangle$). We see that when crossing the longitude at $\phi=\pm\pi/2$ the uncertainty in time for strangeness measurements become ma
ximal compared to $t=0$ measurements ($\equiv$ $\chi^{1/2}$ in the equatorial plan) and again when reaching the south/north pole.
}\label{blochstrange}
\end{figure}

To get a geometrical visualization of the eigenvectors corresponding to the question ``\textit{Is the neutral kaon meson system in the state $|k_n\rangle$ or not at time $t_n$}'' we can represent them by three dimensional real Bloch points
\begin{eqnarray}\label{Blochvector}
\vec{b}^i_n=\langle\chi_n^i|\vec{\sigma}|\chi_n^i\rangle\;.
\end{eqnarray}
on a so called Bloch sphere. The Bloch sphere is an abstract sphere with extreme opposite points (or radial vectors) representing different orthogonal states of a qubit system, i.e. a system with two degrees of freedom. We choose for our representation that the north pole ($\theta=0$) corresponds to the state $|K_S\rangle$ and the state $|K_L\rangle$ ($\theta=\pi$) is presented by the north pole, in case of neglecting $\mathcal{CP}$ violation. Any other pure state of the two--state system corresponds to a point on the surface of the Bloch sphere, in particular the strangeness eigenstates ($|K^0\rangle, (\theta=\frac{\pi}{2},\phi=0)$; $|\bar K^0\rangle, (\theta=\frac{\pi}{2},\phi=\pi)$ are located at the equatorial plane.

In Fig.~\ref{blochstrange} we chose $\theta_n=\pi/2$ and $\phi=0 (\pi)$, i.e. corresponding to the question ``\textit{Is the neutral kaon meson system in the state $|K^0\rangle$ ($|\bar K^0\rangle)$ or not at time $t_n$}''. We draw the eigenvectors of this observable for different time point $t_n$ and observe a spiral movement to the south pole of the Bloch sphere for $\chi^1$ and, consequently, a mirrored movement to the north pole for $\chi^2$ (remember that these two eigenvectors have to be orthogonal which means that $\vec{b}^1=-\vec{b}^2$).

If we are interested in the complementarity for the observable asking the question ``\textit{Is the neutral kaon system in the state $|K^0\rangle$ or not at time $t=0$}'' compared to the question ``\textit{Is the neutral kaon system in the state $|K^0\rangle$ or not at time $t$}'', i.e. comparing the complementary introduced by the time evolution in case of strangeness measurements, we have to consider the scalar product between the radial Bloch vector for $t=0$ (equatorial plane) to a radial Bloch vector corresponding to a later time. We see that the angle given by the scalar product $|\langle\chi_n^1|\chi_m^1\rangle|=|\langle\chi_n^2|\chi_m^2\rangle|$ increases, i.e. the complementary relation increases until the maximal value, then the scalar product between  $|\langle\chi_n^1|\chi_m^2\rangle|=|\langle\chi_n^2|\chi_m^1\rangle|$ becomes greater but due to the fact that the short lived state dies out, the complementarity does not vanish anymore.

In Fig.~\ref{blochstrange} we plotted the eigenvectors of different choices of the azimuth angle $\theta_n$ in case of conservation of the  $\mathcal{CP}$ symmetry $(a),(b)$ and in case of violation $(c),(d)$. We observe that the $|K_L\rangle$ (south pole) and the $|K_S\rangle$ (north pole) are the attractors of the time evolution of the open quantum system for the eigenvectors, respectively, and that the life time observables are persisting in time.  Including $\mathcal{CP}$ violation´(Figs.~(c),(d)) we observe that the lifetime observables are no longer persisting in time, since $\langle K_S|K_L\rangle\not=1$ is no longer orthogonal. Note that, for deriving the physical measurable quantities, i.e. the expectation values we trace over the operator times the initial state which is as well affected by $\mathcal{CP}$ violation. Thus the presented uncertainty is only due to the observable.

\begin{figure}
\centering
(a)\includegraphics[scale=0.33]{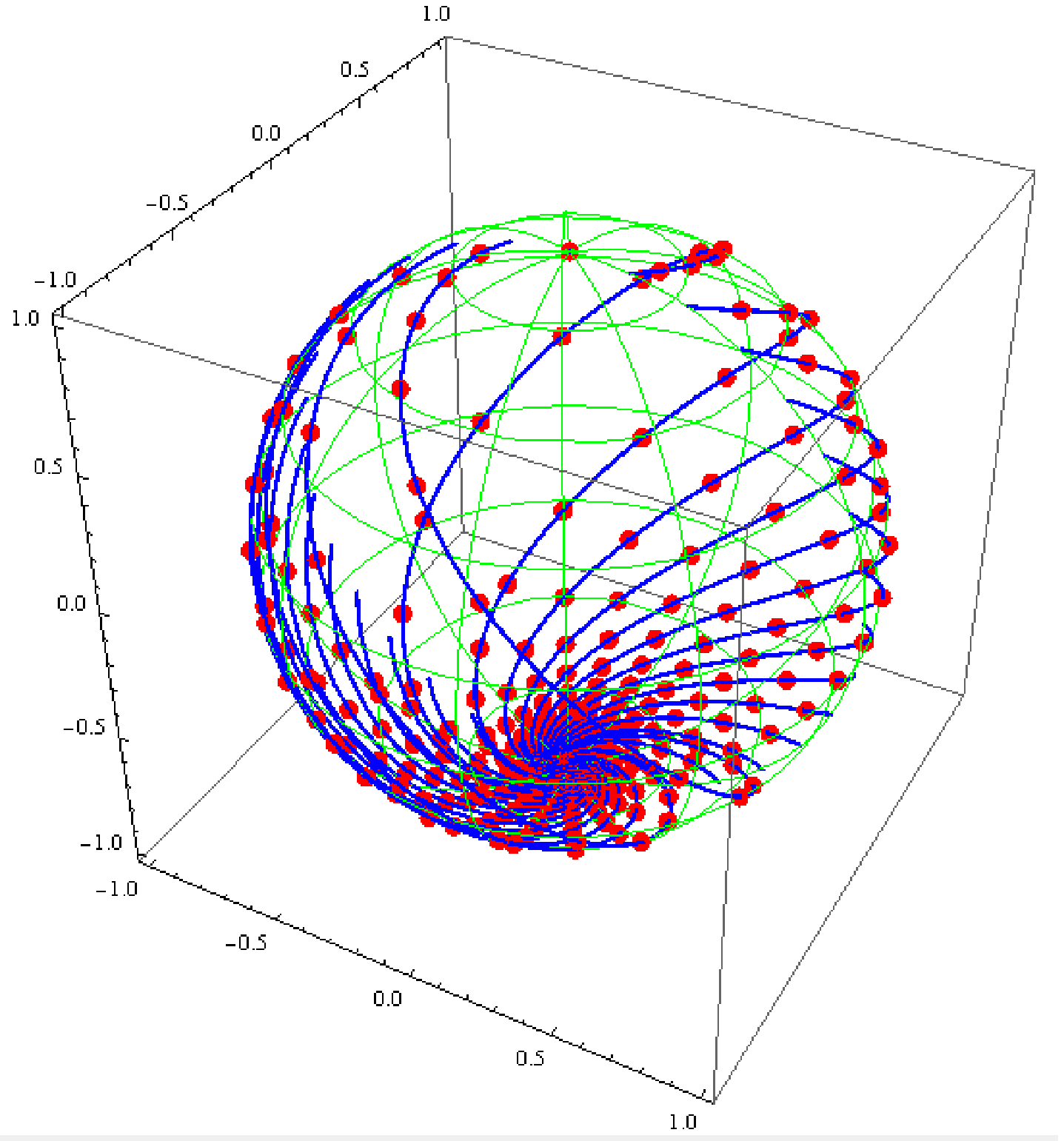}
(b)\includegraphics[scale=0.4]{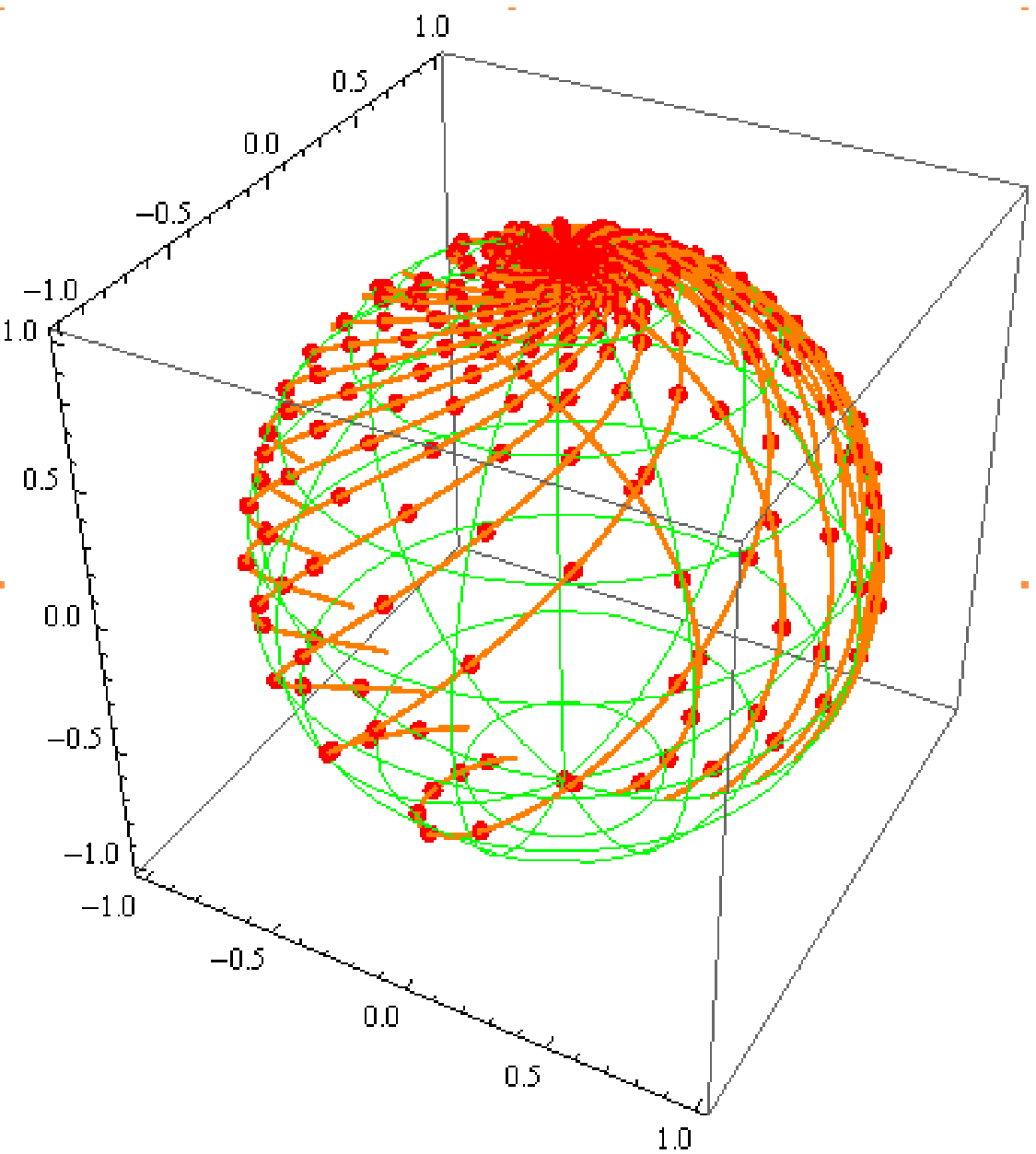}\\
(a)\includegraphics[scale=0.4]{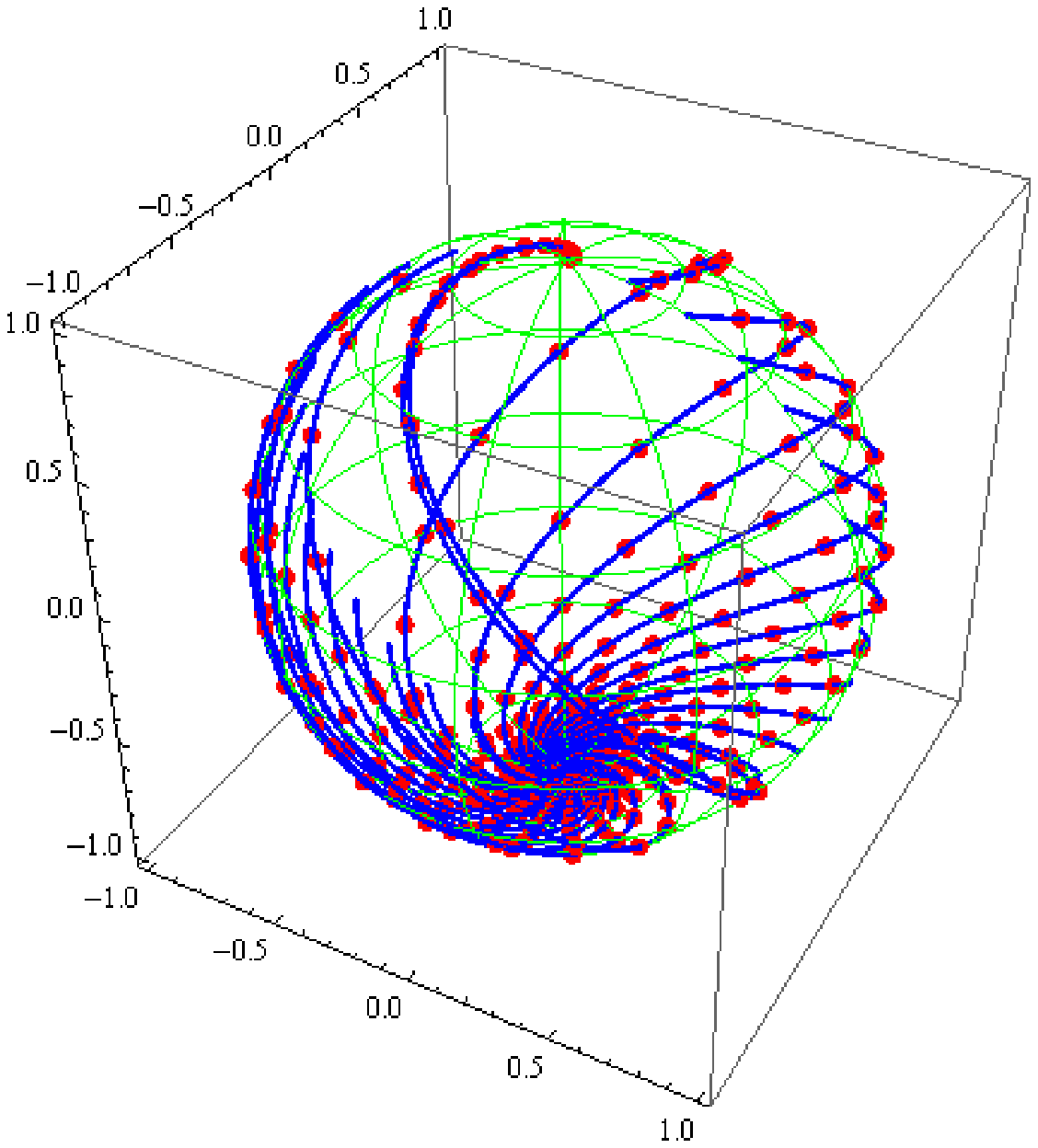}
(b)\includegraphics[scale=0.4]{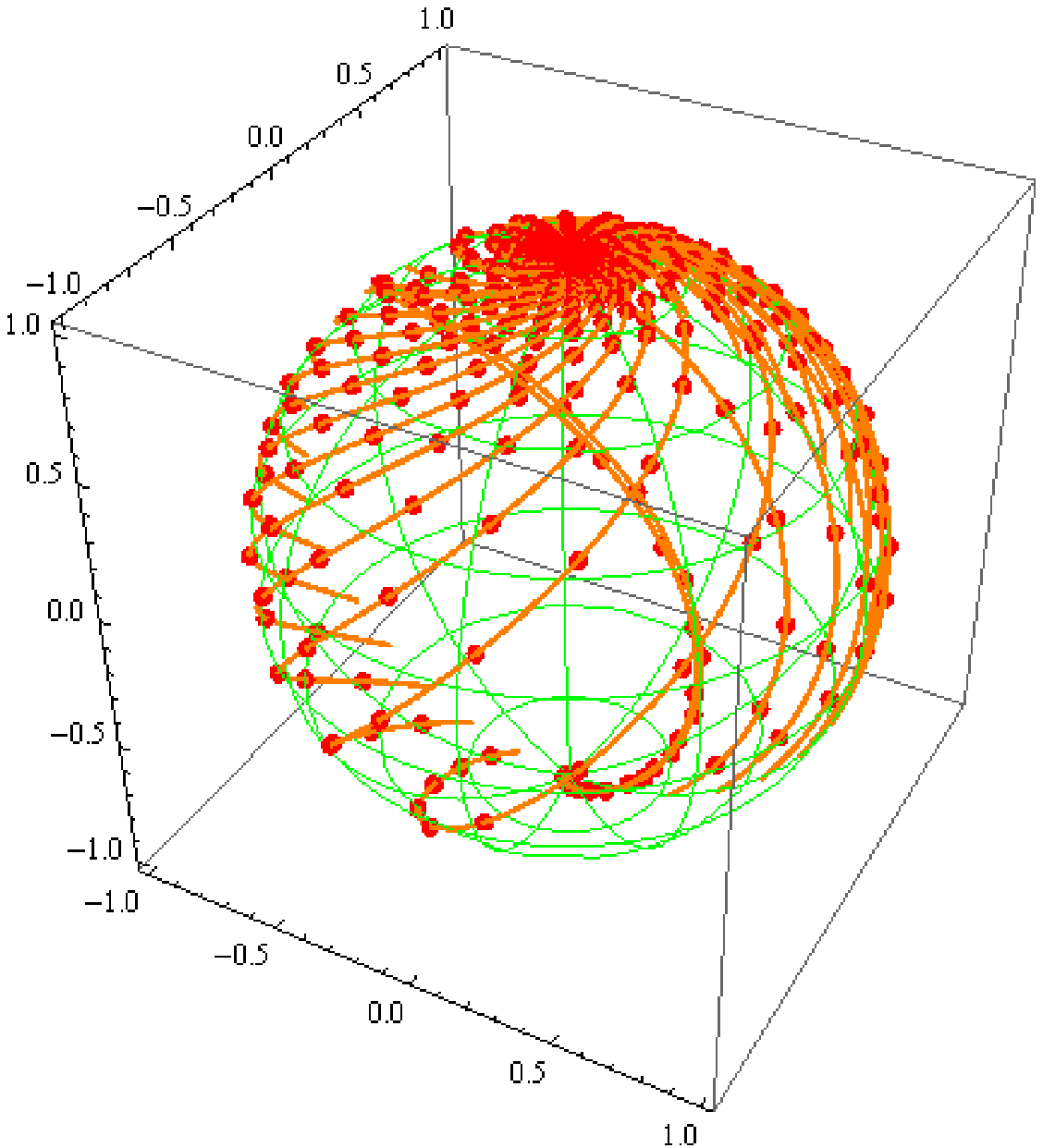}
%(b)\includegraphics[scale=0.5]{blochstrange.eps}
\caption{(Colour online) This picture depicts the time evolution of the eigenstates for different choices of quasispins with $\phi=0,\pi$: (a),(c) of the $\chi^1$'s and (b),(d) of the $\chi^2$'s. We observe that the $|K_L\rangle$ (south pole) and the $|K_S\rangle$ (north pole) are the attractors of the time evolution of the open quantum system for the eigenvectors, respectively. Fig.(c),(d) include also $\mathcal{CP}$ violation. We observe that the eigenvectors of the observable corresponding to the question about a short lived state $K_S$ or not get then also time dependent. Indeed any nonzero value of $\mathcal{CP}$ violating parameter causes a time evolution of the eigenstates. Consequently, a maximal uncertainty is always obtained at a time dependent on the value of the $\mathcal{CP}$ violating parameter. The maximal uncertainty for the $\mathcal{CP}$ violation parameter $\delta=3.322\cdot 10^{-3}$ (world average \cite{PDG}) is reached for $t_n=11.4 \tau_S$ (see Ref.~\cite{Heisikaon}).}\label{blochstrange2}
\end{figure}
%\newpage
%
%
%\begin{figure}
%\centering
%(a)\includegraphics[scale=0.5]{blochkugelstrangea1.eps}
%(b)\includegraphics[scale=0.5]{blochkugelstrangea2.eps}
%(c)\includegraphics[scale=0.5]{blochkugelstrangeb1.eps}
%(d)\includegraphics[scale=0.5]{blochkugelstrangeb2.eps}
%%(b)\includegraphics[scale=0.5]{blochstrange.eps}
%\caption{(Colour online) This picture depicts the time evolution of the eigenstates for different choices of quasispins with $\phi=0,\pi$: (a),(c) of the $\chi^1$'s and (b),(d) of the $\chi^2$'s. We observe that the $|K_L\rangle$ (south pole) and the $|K_S\rangle$ (north pole) are the attractors of the time evolution of the open quantum system for the eigenvectors, respectively. Fig.(c),(d) include also $\mathcal{CP}$ violation. We observe that the eigenvectors of the observable corresponding to the question about a short lived state $K_S$ or not get then also time dependent. Indeed any nonzero value of $\mathcal{CP}$ violating parameter causes a time evolution of the eigenstates. Consequently, a maximal uncertainty is always obtained at a time dependent on the value of the $\mathcal{CP}$ violating parameter. The maximal uncertainty for the $\mathcal{CP}$ violation parameter $\delta=3.322\cdot 10^{-3}$ (world average \cite{PDG}) is reached for $t_n=11.4 \tau_S$ (see Ref.~\cite{Heisikaon}).}\label{blochstrange2}
%\end{figure}

\section{Summary}
We discussed the entropic uncertainty principle for observables in High Energy Physics. We show that strangeness measurements at some time $t_n$ compared to strangeness measurements at a later time $t_m\not= t_n$ introduce always a nonzero uncertainty, i.e a lack of knowledge about the physical properties of this quantum system. Then we discuss the effect of imbalance of matter and antimatter ($CP$ violation) and show that it constrains the potential knowledge an observer can have about the eigenstates of the effective Hamiltonian at a certain time. Given the amount of the $CP$ violation parameter we always find a ``\textit{complementary}'' time, i.e. a time point when the uncertainty reaches the maximal value, i.e. the two observables become complementary.

\ack

I would like to thank the organizers of the DISCETE 2010 conference in Rome, in particular Antonio Di Domenico for inviting and discussions and the Austrian Fund project FWF-P21947N16 and EU project QESSENCE.

\end{document}